\documentclass[acus]{pac99}
\usepackage{epsfig}
\newcommand{\ud}{\mathrm{d}}
\begin{document}
\title{NONLINEAR ACCELERATOR PROBLEMS VIA WAVELETS:\\
1. ORBITAL MOTION IN STORAGE RINGS}
\author{Antonina N.~Fedorova,  Michael G.~Zeitlin, 
IPME, RAS, St.~Petersburg, Russia
\thanks{e-mail: zeitlin@math.ipme.ru}
\thanks{http://www.ipme.ru/zeitlin.html;
        http://www.ipme.nw.ru/zeitlin.html}}
\maketitle

\begin{abstract}
In this series of eight papers  we
present the applications of methods from
wavelet analysis to polynomial approximations for
a number of accelerator physics problems.
In this part,
according to variational approach
 we obtain a representation for orbital particle motion in storage
rings as
a multiresolution (multiscales) expansion in the base of
well-localized in phase space  wavelet basis.
By means of this "wavelet microscope" technique we can take into account
contribution from each scale of resolution.
\end{abstract}
\section{INTRODUCTION}
This is the first part of our eight presentations in which we consider
applications of methods from wavelet analysis to nonlinear accelerator
physics problems. This is a continuation of our results from [1]-[8],
which is based on our approach  to investigation
of nonlinear problems -- general, with additional structures (Hamiltonian,
symplectic or quasicomplex), chaotic, quasiclassical, quantum, which are
considered in the framework of local (nonlinear) Fourier analysis, or wavelet
analysis. Wavelet analysis is a relatively novel set of mathematical
methods, which gives us a possibility to work with well-localized bases in
functional spaces and with the general type of operators (differential,
integral, pseudodifferential) in such bases.
In the parts 1-8 we consider applications
 of wavelet technique  to
nonlinear dynamical problems with polynomial type of nonlinearities.
In this part  we consider this very useful approximation in
the case of orbital motion in storage rings.
Approximation up to octupole terms is only a particular
case of our general construction for n-poles. Our solutions
are parametrized
by solutions of a number of reduced algebraical problems one from which
is nonlinear with the same degree of nonlinearity and the rest  are
the linear problems which correspond to particular
method of calculation of scalar products of functions from wavelet bases
and their derivatives.

\section{Orbital Motion in Storage Rings}

We consider as the main example the particle motion in
storage rings in standard approach, which is based on
consideration in [9].
Starting from Hamiltonian, which described classical dynamics in
storage rings
$
{\cal H}(\vec{r},\vec{P},t)=c\{\pi^2+m_0^2c^2\}^{1/2}+e\phi
$
and using Serret--Frenet parametrization, we have after
standard manipulations with truncation of power series expansion of
square root the following
approximated (up to octupoles) Hamiltonian for orbital motion
in machine coordinates:
\begin{eqnarray}
&&{\cal H}=
   \frac{1}{2}\cdot\frac{[p_x+H\cdot z]^2 + [p_z-H\cdot x]^2}
{[1+f(p_\sigma)]}\nonumber\\
&&+p_\sigma-[1+K_x\cdot x+K_z\cdot z]\cdot f(p_\sigma)\\
&&+\frac{1}{2}\cdot[K_x^2+g]\cdot x^2+\frac{1}{2}\cdot[K_z^2-g]\cdot z^2-
  N\cdot xz \nonumber\\
&&+\frac{\lambda}{6}\cdot(x^3-3xz^2)+\frac{\mu}{24}\cdot(z^4-6x^2z^2+x^4)
\nonumber\\
&&+\frac{1}{\beta_0^2}\cdot\frac{L}{2\pi\cdot h}\cdot\frac{eV(s)}{E_0}\cdot
\cos\left[h\cdot\frac{2\pi}{L}\cdot\sigma+\varphi\right]\nonumber
\end{eqnarray}
Then we use series expansion of function $f(p_\sigma)$ from [9]:
$
f(p_\sigma)=f(0)+f^\prime(0)p_\sigma+f^{\prime\prime}(0)
p_\sigma^2/2+\ldots
=p_\sigma-
p_\sigma^2/(2\gamma_0^2)+\ldots$
and the corresponding expansion of RHS of equations corresponding to (1).
In the following we take into account only an arbitrary
polynomial (in terms of dynamical variables) expressions and
neglecting all nonpolynomial types of expressions, i.e. we
consider such approximations of RHS, which are not more than polynomial
functions in dynamical variables and arbitrary functions of
independent variable $s$ ("time" in our case, if we consider
our system of equations as dynamical problem).

\section{Polynomial Dynamics}

The first main part of our consideration is some variational approach
to this problem, which reduces initial problem to the problem of
solution of functional equations at the first stage and some
algebraical problems at the second stage.
We have the solution in a compactly
supported wavelet basis.
Multiresolution expansion is the second main part of our construction.
The solution is parameterized by solutions of two reduced algebraical
problems, one is nonlinear and the second is some linear
problem, which is obtained from the method of Connection
Coefficients (CC).

\subsection{ Variational Method}
Our problems may be formulated as the systems of ordinary differential
equations
\begin{eqnarray}
{\ud x_i}/{\ud t}=f_i(x_j,t), \quad (i,j=1,...,n)
\end{eqnarray}
with fixed initial conditions $x_i(0)$, where $f_i$ are not more
than polynomial functions of dynamical variables $x_j$
and  have arbitrary dependence of time. Because of time dilation
we can consider  only next time interval: $0\leq t\leq 1$.
 Let us consider a set of
functions
$
 \Phi_i(t)=x_i{\ud y_i}/{\ud t}+f_iy_i
$
and a set of functionals
\begin{eqnarray}
F_i(x)=\int_0^1\Phi_i (t)dt-x_iy_i\mid^1_0,
\end{eqnarray}
where $y_i(t) (y_i(0)=0)$ are dual variables.
It is obvious that the initial system  and the system
\begin{equation}
F_i(x)=0
\end{equation}
are equivalent.
In  the following parts
we consider an approach, which is based on taking into
account underlying symplectic structure and on more useful and flexible
analytical approach, related to bilinear structure of initial functional.
Now we consider formal expansions for $x_i, y_i$:
\begin{eqnarray}\label{eq:pol1}
x_i(t)=x_i(0)+\sum_k\lambda_i^k\varphi_k(t)\quad
y_j(t)=\sum_r \eta_j^r\varphi_r(t),
\end{eqnarray}
where because of initial conditions we need only $\varphi_k(0)=0$.
Then we have the following reduced algebraical system
of equations on the set of unknown coefficients $\lambda_i^k$ of
expansions (\ref{eq:pol1}):
\begin{eqnarray}\label{eq:pol2}
\sum_k\mu_{kr}\lambda^k_i-\gamma_i^r(\lambda_j)=0
\end{eqnarray}
Its coefficients are
\begin{eqnarray}
\mu_{kr}&=&\int_0^1\varphi_k'(t)\varphi_r(t){\rm d}t,\\
\gamma_i^r&=&\int_0^1f_i(x_j,t)\varphi_r(t){\rm d}t.\nonumber
\end{eqnarray}
Now, when we solve system (\ref{eq:pol2}) and determine
 unknown coefficients from formal expansion (\ref{eq:pol1}) we therefore
obtain the solution of our initial problem.
It should be noted if we consider only truncated expansion (\ref{eq:pol1}) with N terms
then we have from (\ref{eq:pol2}) the system of $N\times n$ algebraical equations and
the degree of this algebraical system coincides
 with degree of initial differential system.
So, we have the solution of the initial nonlinear
(polynomial) problem  in the form
\begin{eqnarray}\label{eq:pol3}
x_i(t)=x_i(0)+\sum_{k=1}^N\lambda_i^k X_k(t),
\end{eqnarray}
where coefficients $\lambda_i^k$ are roots of the corresponding
reduced algebraical problem (\ref{eq:pol2}).
Consequently, we have a parametrization of solution of initial problem
by solution of reduced algebraical problem (\ref{eq:pol2}).
The first main problem is a problem of
 computations of coefficients of reduced algebraical
system.
As we will see, these problems may be explicitly solved in wavelet approach.
The obtained solutions are given
in the form (\ref{eq:pol3}),
where
$X_k(t)$ are basis functions and
  $\lambda_k^i$ are roots of reduced
 system of equations.  In our case $X_k(t)$
are obtained via multiresolution expansions and represented by
 compactly supported wavelets and $\lambda_k^i$ are the roots of
corresponding general polynomial  system (\ref{eq:pol2})  with coefficients, which
are given by  CC construction.  According to the
        variational method   to  give the reduction from
differential to algebraical system of equations we need compute
the objects $\gamma ^j_a$ and $\mu_{ji}$,
which are constructed from objects:
\begin{eqnarray}\label{eq:pol4}
\sigma_i&\equiv&\int^1_0X_i(\tau)\ud\tau,\quad
   \nu_{ij}\equiv\int^1_0X_i(\tau)X_j(\tau)\ud\tau,\nonumber\\
    \mu_{ji}&\equiv&\int X'_i(\tau)X_j(\tau)\ud\tau,\\
   \beta_{klj}&\equiv&\int^1_0X_k(\tau)X_l(\tau)X_j(\tau)\ud\tau \nonumber
\end{eqnarray}
for the simplest case of Riccati systems
 (sextupole approximation), where degree of nonlinearity equals to
two. For the general case of arbitrary n we have analogous to (\ref{eq:pol4})
iterated integrals with the degree of monomials in integrand which is one more
bigger than degree of initial system.

\subsection{ Wavelet Computations}
 Now we give construction for
computations of objects (9) in the wavelet case.
We present some details of wavelet machinery in part 2.
We use compactly supported wavelet basis (Fig.~1, for example): orthonormal basis
for functions in $L^2({\bf R})$.
\begin{figure}[ht]
\centering
\epsfig{file=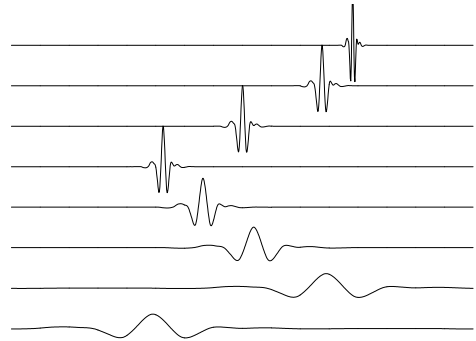, width=82.5mm, bb=0 200 599 590, clip}
\caption{Wavelets at different scales and locations }
\end{figure}

Let be  $ f : {\bf R}\longrightarrow {\bf C}$ and the wavelet
expansion is
\begin{eqnarray}
f(x)=\sum\limits_{\ell\in{\bf Z}}c_\ell\varphi_\ell(x)+
\sum\limits_{j=0}^\infty\sum\limits_{k\in{\bf
Z}}c_{jk}\psi_{jk}(x)
\end{eqnarray}

If in formulae (10) $c_{jk}=0$ for $j\geq J$, then $f(x)$ has an alternative
expansion in terms of dilated scaling functions only
$
f(x)=\sum_{\ell\in{\bf Z}}c_{J\ell}\varphi_{J\ell}(x)
$.
This is a finite wavelet expansion, it can be written solely in
terms of translated scaling functions.
Also we have the shortest possible support: scaling function
$DN$ (where $N$ is even integer) will have support $[0,N-1]$ and
$N/2$ vanishing moments.
There exists $\lambda>0$ such that $DN$ has $\lambda N$
continuous derivatives; for small $N,\lambda\geq 0.55$.
To solve our second associated linear problem we need to
evaluate derivatives of $f(x)$ in terms of $\varphi(x)$.
Let be $
\varphi^n_\ell=\ud^n\varphi_\ell(x)/\ud x^n
$.
We consider computation of the wavelet - Galerkin integrals.
Let $f^d(x)$ be d-derivative of function
 $f(x)$, then we have
$
f^d(x)=\sum_\ell c_l\varphi_\ell^d(x)
$,
and values $\varphi_\ell^d(x)$ can be expanded in terms of
$\varphi(x)$
\begin{eqnarray}
\phi_\ell^d(x)=\sum\limits_m\lambda_m\varphi_m(x),
 \end{eqnarray}
where $\lambda_m=\int\varphi_\ell^d(x)\varphi_m(x)\ud x$ are wavelet-Galerkin integrals.
The coefficients $\lambda_m$  are 2-term connection
coefficients. In general we need to find $(d_i\geq 0)$
\begin{eqnarray}
\Lambda^{d_1 d_2 ...d_n}_{\ell_1 \ell_2 ...\ell_n}=
 \int\limits_{-\infty}^{\infty}\prod\varphi^{d_i}_{\ell_i}(x)dx
\end{eqnarray}
For Riccati case (sextupole) we need to evaluate two and three
connection coefficients
\begin{eqnarray}
\Lambda_\ell^{d_1
d_2}&=&\int^\infty_{-\infty}\varphi^{d_1}(x)\varphi_\ell^{d_2}(x)dx,
\\
\Lambda^{d_1 d_2
d_3}&=&\int\limits_{-\infty}^\infty\varphi^{d_1}(x)\varphi_
\ell^{d_2}(x)\varphi^{d_3}_m(x)dx\nonumber
\end{eqnarray}
According to CC method [10] we use the next construction. When $N$  in
scaling equation is a finite even positive integer the function
$\varphi(x)$  has compact support contained in $[0,N-1]$.
For a fixed triple $(d_1,d_2,d_3)$ only some  $\Lambda_{\ell
 m}^{d_1 d_2 d_3}$ are nonzero: $2-N\leq \ell\leq N-2,\quad
2-N\leq m\leq N-2,\quad |\ell-m|\leq N-2$. There are
$M=3N^2-9N+7$ such pairs $(\ell,m)$. Let $\Lambda^{d_1 d_2 d_3}$
be an M-vector, whose components are numbers $\Lambda^{d_1 d_2
d_3}_{\ell m}$. Then we have the first reduced algebraical system
: $\Lambda$
satisfy the system of equations $(d=d_1+d_2+d_3)$
\begin{eqnarray}
A\Lambda^{d_1 d_2 d_3}&=&2^{1-d}\Lambda^{d_1 d_2 d_3},
\\
A_{\ell,m;q,r}&=&\sum_p a_p a_{q-2\ell+p}a_{r-2m+p}.\nonumber
\end{eqnarray}
By moment equations we have created a system of $M+d+1$
equations in $M$ unknowns. It has rank $M$ and we can obtain
unique solution by combination of LU decomposition and QR
algorithm.
The second  reduced algebraical system gives us the 2-term connection
coefficients.
For nonquadratic case we have analogously additional linear problems for
objects (12).
Solving these linear problems we obtain the coefficients of nonlinear
algebraical system (6) and after that we obtain the coefficients of wavelet
expansion (8). As a result we obtained the explicit time solution  of our
problem in the base of compactly supported wavelets with
the best possible localization in the phase space, which allows
us to control contribution from each scale of underlying multiresolution
expansions.

In the following parts we consider extension of this approach to the case of
(periodic)
boundary conditions, the case of presence of arbitrary variable coefficients
and more flexible biorthogonal wavelet approach.

We are very grateful to M.~Cornacchia (SLAC),
W.~Her\-r\-man\-nsfeldt (SLAC),
Mrs. J.~Kono (LBL) and
M.~Laraneta (UCLA) for
 their permanent encouragement.

 \end{document}